%started on May 31, 1997
%final version on June 20, 1997
%
%     This is a contribution to the special issue in 
%                Foundation of Physics
%     dedicated to Prof. Namiki's 70th birthday 
%
%     TIME DEVELOPMENT OF A WAVE PACKET AND THE TIME DELAY
%\input ../xiipt \xiipoint
%\jfont\xiimin=min10 %scaled \magstep1
%\jfont\xiigt=goth10
\font\xiirm=cmr12       \font\ninerm=cmr9
\font\xiii=cmmi12       \font\ninei=cmmi9
 \skewchar\xiii='177
\font\xiisy=cmsy10 scaled \magstep1
 \skewchar\xiisy='60    \font\ninesy=cmsy9
\font\tenex=cmex10
\font\xiiti=cmti12
\font\xiibf=cmbx12

                        \font\ninebf=cmbx9

\def\xiipoint{\def\rm{\fam0\xiirm}%
  \textfont0=\xiirm    \scriptfont0=\ninerm   \scriptscriptfont0=\sevenrm
  \textfont1=\xiii     \scriptfont1=\ninei    \scriptscriptfont1=\seveni
  \textfont2=\xiisy    \scriptfont2=\ninesy   \scriptscriptfont2=\sevensy
  \textfont3=\tenex    \scriptfont3=\tenex    \scriptscriptfont3=\tenex
  \textfont\itfam=\xiiti                      \def\it{\fam\itfam\xiiti}%
  \textfont\bffam=\xiibf                      \scriptfont\bffam=\ninebf
  \scriptscriptfont\bffam=\sevenbf            \def\bf{\fam\bffam\xiibf}%
  \normalbaselineskip=16.8pt
  \setbox\strutbox=\hbox{\vrule height11.9pt depth4.9pt width0pt}%
  \normalbaselines\rm}%\xiimin}
\abovedisplayskip=14pt plus 3.6pt minus 10.8pt
\abovedisplayshortskip=0pt plus 3.6pt
\belowdisplayskip=\abovedisplayskip
\belowdisplayshortskip=8.4pt plus 3.6pt minus 4.8pt
\footline={\hss\xiirm\folio\hss}
\xiipoint
%
%-----MACROS \gsim, \lsim, \scriptgsim and \scriptlsim
%            for mathematical symbols both in the text-, displayed- and 
%            scriptstyles
%
\def\underbuildrel#1\over#2{\mathrel{\mathop{\kern0pt #1}\limits_{#2}}}
\def\originalgsim{\underbuildrel>\over\sim}
\def\originallsim{\underbuildrel<\over\sim}
\def\gsim{\mathrel{\raise 2pt\hbox{$\textstyle \originalgsim$}}}
\def\lsim{\mathrel{\raise 2pt\hbox{$\textstyle \originallsim$}}}
\def\scriptgsim{\mathrel{\raise 1pt\hbox{$\scriptscriptstyle \originalgsim$}}}
\def\scriptlsim{\mathrel{\raise 1pt\hbox{$\scriptscriptstyle \originallsim$}}}
\def\ppsi#1{\langle p|\psi\rangle_{#1}}
\def\puppsi#1{\langle p\uparrow|\psi\rangle_{#1}}
\def\pdownpsi#1{\langle p\downarrow|\psi\rangle_{#1}}
\def\qpsi#1{\langle q|\psi\rangle_{#1}}
\def\quppsi#1{\langle q\uparrow|\psi\rangle_{#1}}
\def\qdownpsi#1{\langle q\downarrow|\psi\rangle_{#1}}
\def\Cup#1{C_\uparrow(#1)}
\def\Cdown#1{C_\downarrow(#1)}
\def\tCup#1{\tilde{C}_\uparrow(#1)}
\def\tCdown#1{\tilde{C}_\downarrow(#1)}
%
%------------------------------------------------------------------------------
%
\rightline{June 1997}
\rightline{WU-HEP-1997-1}
\vskip2.5cm
\centerline{\bf Time development of a wave packet and the time delay}
\vskip1.5cm
\centerline{Hiromichi Nakazato}
\vskip1cm
\centerline{\it Department of Physics, Waseda University}
\centerline{\it Okubo 3-4-1, Shinjuku, Tokyo 169, Japan}
\vfill
\centerline{\bf Abstract}
\medskip
      A one-dimensional scattering problem off a $\delta$-shaped potential is solved analytically and the time development of a wave packet is derived from the time-dependent Schr\"odinger equation.
The exact and explicit expression of the scattered wave packet supplies us with interesting information about the ``time delay" by potential scattering in the asymptotic region.
It is demonstrated that a wave packet scattered by a spin-flipping potential can give us quite a different value for the delay times from that obtained without spin-degrees of freedom.
\vskip 60pt
\eject
%------------------------------------------------------------------------------
\centerline{\bf \S 1.  Introduction and summary}
%------------------------------------------------------------------------------
\medskip
     It is well known that the ``time" occupies a rather peculiar position in quantum mechanics in the sense that it is just a parameter in the Schr\"odinger equation and no satisfactory ``time operator" [1] 
has been found or devised so far, even though it is certainly measurable in experiments.
This peculiarity can be one of the main sources of controversies over, for example, the definition of the so-called ``tunneling time" 
%[\ref:tun-time]
%\item{[\ref:tun-time]} 
or the interaction time of a particle with a potential when the particle is represented by an almost monochromatic wave packet and its spatial width is longer than the range of the potential.

To treat this kind of problem, one sometimes uses the {\it time-independent\/} Schr\"odinger equation and forms a wave packet by superposing its solutions.
Even though it can supply us with some information on the time development of the system under consideration, one has, at the same time, to remember in which sense this kind of treatment can represent the actual dynamical process:
The solution to the time-independent Schr\"odinger equation is considered to describe the actual scattering process as an approximately stationary process, developing around the scatterer, where both the incoming and outgoing waves are present.
The use of stationary solutions is justified in this restricted sense.
Of course, we may construct a wave packet from these solutions with an appropriate weight function, in order to see the explicit time dependence.
In this case, however, we have to be careful about the choice of the very moment $t=0$ since the actual scattering process is by no means translationally invariant (i.e.\ there is a moment when the particle is injected).
In connection to this, we also need to correctly understand the meaning of the weight function.
In this respect, it would be more natural and can be unambiguous to introduce a wave packet from the beginning to represent the incident particle and to examine its time evolution on the basis of the {\it time-dependent\/} Schr\"odinger equation.
We shall set $t=0$ when the particle is injected and the weight function is so chosen as to reproduce the incident wave packet.

     The purpose of this paper is twofold.

     First, a one-dimensional time-dependent Schr\"o\-ding\-er equation for a particle scattered by a simple $\delta$-shaped potential is solved analytically to give an exact and explicit wave-packet solution in \S2.
Even though it remains at an elementary level, the derivation turns out to be rather instructive and suggestive.
The exact expression of the wave packet enables us to understand how its asymptotic behavior emerges and is to be restricted on the energy shell as $t\to\infty$.
The connection to the stationary solutions is easily seen.
The analysis is extended to incorporate possible internal degrees of freedom in \S3.
As a concrete and still solvable example, the scattering of a spin-1/2 particle off a spin-flipping $\delta$-potential is considered and its wave-packet solution is derived.

     Second, in \S4., we choose an almost monochromatic incident wave packet (i.e.\ spatially broad wave packet) and derive its asymptotic form, from which the mean position is calculated and compared with that of a free wave packet without interaction.
In the simplest case with no internal degrees of freedom, it turns out that in spite of the vanishing interaction range of the $\delta$-potential, the scattered wave packet exhibits a finite time delay or advance, depending on whether the interaction is repulsive or attractive.
This delay (or advance) time is shown to agree with that derived from the energy derivative of the phase shift [2]
caused by the potential scattering.
In the spin-1/2 case, there are four scattering channels depending on whether the particle is transmitted or reflected and the spin is flipped or not.
It is shown explicitly that contrary to the first case without spin-degrees of freedom, there remains no time delay or advance in any channels.
This can be traced back to the reality of the scattering-matrix elements (i.e.\ the transmission and reflection coefficients) in this case.
Notice that these two simple examples and their results imply that the notion of time delay of a wave packet can be quite different from the classical counterpart $\sim$ [interaction range]/[particle's velocity] and its estimation fully needs quantum mechanical treatment.

\bigskip
%------------------------------------------------------------------------------
\centerline{\bf\S2.~Wave-packet solution for scattering off a $\delta$-potential}
%------------------------------------------------------------------------------
\medskip
     Let us consider a simple scattering problem, i.e.\ a one-dimensional scattering of a particle off a single $\delta$-potential.
The Hamiltonian is given by
$$
H={p^2\over2m}+g\delta(x).
\eqno(2.1)
$$
We shall solve the time-dependent Schr\"odinger equation for a wave function in momentum space $\ppsi t$,
$$
i\hbar{\partial\over\partial t}\ppsi t
=\langle p|H|\psi\rangle_t
={p^2\over2m}\ppsi t+gC(t),
\eqno(2.2)
$$
where we have put
$$
C(t)\equiv\int{dq\over2\pi\hbar}\qpsi t.
\eqno(2.3)
$$
The solution is easily found
$$
\ppsi t=e^{-ip^2t/2m\hbar}\ppsi0
        -i{g\over\hbar}\int^t_0dt'e^{-ip^2(t-t')/2m\hbar}C(t')
\eqno(2.4)
$$
in terms of the initial wave function $\ppsi0$ and $C(t)$, the latter of which satisfies the consistency condition
$$
C(t)=\int{dq\over2\pi\hbar}e^{-iq^2t/2m\hbar}\qpsi0
     -i{g\over\hbar}\int{dq\over2\pi\hbar}\int^t_0dt'
       e^{-iq^2(t-t')/2m\hbar}C(t').
\eqno(2.5)
$$
This equation can be reduced to an algebraic one by means of the Laplace transform
$$
\tilde C(s)=\int^\infty_0dte^{-st}C(t),\qquad(s>0).
\eqno(2.6)
$$
The solution $\tilde C(s)$ is
$$ 
\tilde C(s)
={1\over1+i(g/\hbar){\cal F}(s)}
 \int{dq\over2\pi\hbar}{\qpsi0\over s+iq^2/2m\hbar},
\eqno(2.7)
$$
where
$$
{\cal F}(s)=\int{dq\over2\pi\hbar}{1\over s+iq^2/2m\hbar}.
\eqno(2.8)
$$
In order to obtain $C(t)$ from $\tilde C(s)$ via the inverse Laplace transformation, we need to make an analytic continuation of the latter from the original domain of $s>0$ into the left-half complex $s$-plane where ${\rm Re}\, s<0$.
Since the above function $\cal F$ is easily evaluated to be 
$$
{\cal F}(s)=\sqrt{m\over2\hbar}{1\over\sqrt{is}},
\eqno(2.9)
$$
where $\sqrt s$ has a cut along the negative imaginary $s$-axis, we can write down $C(t')$ ($t'>0$) as
$$\eqalign{
C(t')&=\int{ds\over2\pi i}e^{st'}\tilde C(s)\cr
     &=\int{dq\over2\pi\hbar}\qpsi0
       \int_{-i\infty+\epsilon}^{i\infty+\epsilon}{ds\over2\pi i}
       {\sqrt{is}\{\sqrt{is}-i(g/\hbar)\sqrt{m/2\hbar}\}\,e^{st'}
        \over
        i(s-img^2/2\hbar^3)(s+iq^2/2m\hbar)}.\cr}
\eqno(2.10)
$$

     Observe that the above integrand in (2.10) has a cut along the negative imaginary $s$-axis and two simple poles at $s=-iq^2/2m\hbar$ (for fixed $q$) and $s=img^2/2\hbar^3$, the latter of which exists only on the second Riemannian sheet if the coupling constant $g$ is positive (repulsive potential) and on the first sheet if $g<0$ (attractive potential).
Therefore the original contour shall be deformed, avoiding these simple poles, to encircle in the left-half plane of the second Riemannian sheet.
See Fig.~1.  
\bigskip
\centerline{
\vbox{\hsize2.5cm
      \hrule
      \smallskip
      \centerline{Fig.~1.}
      \smallskip
      \hrule}
}
\bigskip\noindent
The choice of these deformed contours facilitates the integration over $s$ and we obtain
$$\eqalign{
C(t')&=\int{dq\over2\pi\hbar}\qpsi0\cr
     &\qquad\times\Biggl[
       {1\over1+img/\hbar|q|}e^{-iq^2t'/2m\hbar}
        +{1\over1+(\hbar q/mg)^2}e^{img^2t'/2\hbar^3}\cr
     &\phantom{\qquad\times\Biggl[}\qquad
        +{g\over\pi\hbar}\sqrt{m\over2\hbar}{\cal P}
          \int_0^\infty dy{\sqrt ye^{iyt'}
                           \over(y-mg^2/2\hbar^3)(y+q^2/2m\hbar)}
     \Biggr],\cr
}
\eqno(2.11)
$$
which is valid irrespectively of the sign of $g$.
Each term in the square parentheses in (2.11) corresponds to the contribution arising from each singularity of the integrand mentioned above.

     This explicit expression for $C(t')$ leads us to an exact expression%
\footnote{$^*$}
{It is possible to get an apparently different expression for $\ppsi t$, if a different contour is chosen in (2.10), even though it is expressing completely the same quantity.  
The final form is dependent on the choice of the contour and one may choose a convenient contour, according to one's purpose.}
of the wave function in $p$-space [see Eq.~(2.4)]
$$\eqalignno{
\ppsi t
&=e^{-ip^2t/2m\hbar}\cr
&\quad\times\Biggl[
  \ppsi0
  -i{g\over\hbar}\int{dq\over2\pi\hbar}\langle q|\psi\rangle_0
  \Biggl\{
  {1\over1+img/\hbar|q|}\Delta_t[(p^2-q^2)/2m\hbar]\cr
&\phantom{\quad\times\Biggl[\ppsi0-i{g\over\hbar}
          \int{dq\over2\pi\hbar}\langle q|\psi\rangle_0}
 +{1\over1+(\hbar q/mg)^2}\Delta_t[(p^2+(mg/\hbar)^2)/2m\hbar]\cr
&\phantom{\quad\times\Biggl[\ppsi0-i{g\over\hbar}
          \int{dq\over2\pi\hbar}\langle q|\psi\rangle_0}
 +{g\over\pi\hbar}\sqrt{m\over2\hbar}{\cal P}
   \int_0^\infty dy
   {\sqrt y\,\Delta_t[y+p^2/2m\hbar]\over(y-mg^2/2\hbar^3)(y+q^2/2m\hbar)}
       \Biggr\}\Biggr],\cr
&&(2.12)\cr}
$$
where
$$
\Delta_t[x]\equiv{e^{ixt}-1\over ix}.
\eqno(2.13)
$$
Notice that this is an exact expression [3] for the wave function at $t>0$ and no assumption has been made in its derivation.

     Obviously the first term represents the free evolution of the incident wave packet, while the other terms stand for the scattered components.
It would be interesting to observe that apart from the factor $\exp(-ip^2t/2m\hbar)$, the $t$ dependence appears only through the functions $\Delta_t$s, each of which contains an energy-nonconserving component.
The appearance of these energy-nonconserving terms is due to the energy-time uncertainty relation and they are expected to die out when the wave packet has passed through the interaction region and reaches the so-called wave zone.
In order to understand this phenomenon clearly, let us consider the asymptotic limit $t\to\infty$ in (2.12).
Notice that for large $t$, the function $\Delta_t[x]$ approaches the Dirac $\delta$-function
$$
\Delta_t[x]\buildrel t\to\infty\over\longrightarrow2\pi\delta(x).
\eqno(2.14)
$$
Therefore in this asymptotic limit, only one of three $\Delta_t$s, that is $\Delta_t[(p^2-q^2)/2m\hbar]$ survives to give the energy conservation $p^2/2m=q^2/2m$, while the others $\Delta_t[(p^2+(mg/\hbar)^2)/2m\hbar]$ and $\Delta_t[y+p^2/2m\hbar]$ vanish since their arguments are positive definite.
It is worth noticing that the argument of $\Delta_t$ is nothing but the sum of the initial energy and the imaginary part of the location of the simple pole in the $s$-plane so that contributions coming from poles in the upper-half complex $s$-plane are to die out in the $t\to\infty$ limit, since they have positive imaginary parts and are not able to satisfy the energy conservation:
Only those singularities located on the lower-half $s$-plane have relevance in the asymptotic limit.

     It is now clear that in the asymptotic region $t\to\infty$, the wave packet $\ppsi t$ is confined to the energy shell and is given by 
$$\eqalignno{
\ppsi t\buildrel t\to\infty\over\longrightarrow
&\,e^{-ip^2t/2m\hbar}\left\{
   \ppsi0
   -i{mg\over\hbar|p|}{1\over1+img/\hbar|p|}
    \Bigl(\ppsi0+\langle-p|\psi\rangle_0\Bigr)\right\}\cr
=&\,e^{-ip^2t/2m\hbar}\left\{
   {1\over1+img/\hbar|p|}\ppsi0
   +{-img/\hbar|p|\over1+img/\hbar|p|}\langle-p|\psi\rangle_0
  \right\}.&(2.15)\cr
}
$$
Observe that the two terms in the last expression correspond to the transmitted and reflected waves and the former is given by the sum of the free (nonscattered) and the scattered waves.
Equation (2.15) shows clearly and explicitly the connection of the wave-packet (i.e.\ the time-dependent) solution to the stationary solution in the asymptotic region:
In fact, the two factors in front of $\ppsi0$ and $\langle-p|\psi\rangle_0$ are nothing but the transmission ($\cal T$) and reflection ($\cal R$) coefficients, respectively, obtained for plane-wave scattering with definite momentum $p$
$$
{\cal T}(p)={1\over1+img/\hbar|p|},\qquad
{\cal R}(p)={-img/\hbar|p|\over1+img/\hbar|p|}.
\eqno(2.16)
$$

\bigskip
%-----------------------------------------------------------------------------
\centerline{\bf\S3.~Scattering off a spin-flipping $\delta$-potential 
                    and its wave-packet solution}
%-----------------------------------------------------------------------------
\medskip
     In this section, we consider a scattering of a particle endowed with internal degrees of freedom from a $\delta$-shaped potential which is able to change the internal state of the particle.
For definiteness and simplicity, let a spin-1/2 particle prepared in the up-state be injected at $t=0$ to the spin-flipping potential.
Two spin-eigenstates of the particle, energetically separated by $\hbar\omega$, shall be taken to be those of the third Pauli matrix $\sigma_3$, so that the total Hamiltonian is
$$
H={p^2\over2m}+{\hbar\omega\over2}(1+\sigma_3)+g\delta(x)\sigma_1.
\eqno(3.1)
$$
The last term makes the particle's spin flip when it interacts with the potential.

     It is important to note that here we retain the spatial degrees of freedom of the particle.
This is in contrast to the case in which, e.g.\ spin flip by a magnetic field is considered, where the spatial degrees of freedom are completely neglected and the interaction time is set to be [range of magnetic field]/[particle's velocity] from the outset.
This is justifiable under the situation where the range of interaction is much larger than the width of the wave packet representing the incident particle, even if the latter is broad enough to approximate the particle as a plane wave.
Obviously this condition does not hold in our case, since the interaction range of the $\delta$-potential is 0.
Actually, a question about the interaction time for such a short-range potential is one of the motivations of the present work.

     If we decompose the wave function in terms of momentum and spin eigenstates
$$
|\psi\rangle_t=\int dp\left[|p\uparrow\rangle\puppsi t
                           +|p\downarrow\rangle\pdownpsi t\right],
\eqno(3.2)
$$
the Schr\"odinger equation becomes the coupled equations between $\puppsi t$ and $\pdownpsi t$
$$\eqalignno{
i\hbar{\partial\over\partial t}\puppsi t
&=\Bigl({p^2\over2m}+\hbar\omega\Bigr)\puppsi t+g\Cdown t,&\rm(3.3a)\cr
i\hbar{\partial\over\partial t}\pdownpsi t
&={p^2\over2m}\pdownpsi t+g\Cup t,&\rm(3.3b)\cr}
$$
where
$$
\Cup t\equiv\int{dq\over2\pi\hbar}\quppsi t,\qquad
\Cdown t\equiv\int{dq\over2\pi\hbar}\qdownpsi t.
\eqno(3.4)
$$
Since the initial state is set in a spin-up state, $\pdownpsi0=0$ and the solution to Eqs.~(3.3) is given by
$$\displaylines{
\quad\puppsi t=e^{-i(p^2/2m\hbar+\omega)t}\puppsi0
               -i{g\over\hbar}
               \int_0^tdt'\,e^{-i(p^2/2m\hbar+\omega)(t-t')}\Cdown{t'},
\hfill\rm(3.5a)\cr
\noalign{\smallskip}
\quad\pdownpsi t=-i{g\over\hbar}
                 \int_0^tdt'\,e^{-ip^2(t-t')/2m\hbar}\Cup{t'}.
\hfill\rm(3.5b)\cr}
$$
The consistency conditions that the above functions $\Cup t$ and $\Cdown t$ should satisfy
$$\displaylines{
\quad\Cup t
=\int{dq\over2\pi\hbar}\,e^{-i(q^2/2m\hbar+\omega)t}\quppsi0
  -i{g\over\hbar}\int{dq\over2\pi\hbar}\int_0^tdt'\,
   e^{-i(q^2/2m\hbar+\omega)(t-t')}\Cdown{t'},\hfill\rm(3.6a)\cr
\noalign{\medskip}
\quad\Cdown t
=-i{g\over\hbar}\int{dq\over2\pi\hbar}\int_0^tdt'\,
   e^{-iq^2(t-t')/2m\hbar}\Cup{t'}\hfill\rm(3.6b)\cr}
$$
are converted to algebraic equations after the Laplace transformation.
After an elementary calculation we find the solution
$$
\left[\matrix{\tCup s  \cr
\noalign{\smallskip}
              \tCdown s\cr}\right]
=\int{dq\over2\pi\hbar}{\quppsi0\over s+i(q^2/2m\hbar+\omega)}
 {\sqrt{s(s+i\omega)}\over\sqrt{s(s+i\omega)}-(img^2/2\hbar^3)}
 \left[\matrix{1                                   \cr
\noalign{\smallskip}
               -(g/\hbar)\sqrt{im/2\hbar s}\cr}\right].
\eqno(3.7)
$$
It is clear that the Laplace transforms $\tCup s$ and $\tCdown s$ have two simple poles at $s=i(\sqrt{\omega^2+(mg^2/\hbar^3)^2}-\omega)/2$ and $s=-i(q^2/2m\hbar+\omega)$, two branch points at $s=0$ and $s=-i\omega$ and a cut along the negative imaginary $s$-axis.
The final expression of $\Cup{t'}$ and $\Cdown{t'}$ is dependent on how the integration contour is deformed in the complex $s$-plane. 

     For the evaluation of $\Cdown{t'}$, we shall deform the contour as in Fig.~2(a)
\bigskip\medskip
\centerline{
\vbox{\hsize2.5cm
      \hrule
      \smallskip
      \centerline{Fig.~2.}
      \smallskip
      \hrule}
}
\bigskip\noindent
to obtain
$$\eqalign{
\Cdown{t'}&=\int{dq\over2\pi\hbar}\langle q\uparrow|\psi\rangle_0\cr
          &\quad\times\Biggl[
           -i{mg\over\hbar}e^{-i(q^2/2m\hbar+\omega)t'}
           {\sqrt{q^2(q^2+2m\hbar\omega)}
            \over
            \sqrt{q^2(q^2+2m\hbar\omega)}+(mg/\hbar)^2}\cr
          &\phantom{\quad\times\Biggl[}
           +i{g\over\hbar}\sqrt{m\over2\hbar}{\cal P}
           \int_0^\infty{dy\over\pi}{e^{iyt'}\over y+(q^2/2m\hbar+\omega)}
           {\sqrt{y+\omega}\over\sqrt{y(y+\omega)}-(mg^2/2\hbar^3)}\cr
          &\phantom{\quad\times\Biggl[}
           +i\sqrt{m^3\over2\hbar^3}\Bigl({g\over\hbar}\Bigr)^3
           \int_0^\omega{dy\over2\pi}{e^{-iyt'}\over y-(q^2/2m\hbar+\omega)}
           {\sqrt{\omega-y}\over y(\omega-y)+(mg^2/2\hbar^3)^2}\Biggr].\cr}
\eqno(3.8)
$$
Inserting this into (3.5a) and integrating over $t'$, we reach an exact formula for $\puppsi t$.
In particular, in the asymptotic region $t\to\infty$, the first, second and third terms in the above square parentheses give contributions proportional to $\delta(p^2-q^2)$, $\delta(y+p^2/2m\hbar+\omega)$ and $\delta(y-p^2/2m\hbar-\omega)$, respectively, after the $t'$-integration, so that only the first term can survive to give the asymptotic behavior of  $\puppsi t$
$$\eqalignno{
\puppsi t\buildrel{t\to\infty}\over\longrightarrow
&\,e^{-i(p^2/2m\hbar+\omega)t}\puppsi0\cr
&\qquad
 +{mg^2\over\hbar^2}\int{dq\over2\pi\hbar}\langle q\uparrow|\psi\rangle_0
  {\sqrt{q^2}\,e^{-i(p^2/2m\hbar+\omega)t}\over
   \sqrt{q^2(q^2+2m\hbar\omega)}+(mg/\hbar)^2}
  4\pi m\hbar\delta(p^2-q^2)\cr
=&\,e^{-i(p^2/2m\hbar+\omega)t}\Biggl[
  {\sqrt{p^2(p^2+2m\hbar\omega)}\over
   \sqrt{p^2(p^2+2m\hbar\omega)}+(mg/\hbar)^2}\puppsi0\cr
&\phantom{e^{-i(p^2/2m\hbar+\omega)t}\Biggl[}\qquad
  +{-(mg/\hbar)^2\over\sqrt{p^2(p^2+2m\hbar\omega)}+(mg/\hbar)^2}
  \langle-p\uparrow|\psi\rangle_0\Biggr].&(3.9)\cr}
$$
The first and second terms in the parentheses respectively stand for the components, transmitted and reflected through the potential without changing the spin state and the factors multiplying them coincide with the ordinary transmission ($\cal T_\uparrow$) and reflection ($\cal R_\uparrow$) coefficients obtained by solving the time-independent Schr\"odinger equation.
Observe that these coefficients are both real quantities.

     Similar treatment can be done for the evaluation of $\Cup{t'}$:
We shall take the deformed contour, depicted in Fig.~2(b), and get
$$\eqalignno{
\Cup{t'}=\int{dq\over2\pi\hbar}\quppsi0\Biggl[
         &{q^2(q^2+2m\hbar\omega)\over q^2(q^2+2m\hbar\omega)-(mg/\hbar)^4}
          e^{-i(q^2/2m\hbar+\omega)t'}\cr
         &-i{mg^2\over\hbar^3}{\cal P}\int_\omega^\infty{dy\over2\pi}
          {e^{-iyt'}\over y-(q^2/2m\hbar+\omega)}
          {\sqrt{y(y-\omega)}\over y(y-\omega)-(mg^2/2\hbar^3)^2}\Biggr].\cr
&&(3.10)\cr}
$$
In the $t\to\infty$ limit, $\pdownpsi t$ is shown to behave like
$$\eqalign{
\pdownpsi t\buildrel t\to\infty\over\longrightarrow
&-i{mg\over\hbar}\theta\Bigl[{p^2\over2m}-\hbar\omega\Bigr]
 e^{-ip^2t/2m\hbar}\cr
&\quad\times\Biggl[
 {\sqrt{p^2}\over\sqrt{p^2(p^2-2m\hbar\omega)}+(mg/\hbar)^2}
 \langle\sqrt{p^2-2m\hbar\omega}\uparrow|\psi\rangle_0\cr
&\quad\phantom{\times\Biggl[}
 +{\sqrt{p^2}\over\sqrt{p^2(p^2-2m\hbar\omega)}+(mg/\hbar)^2}
 \langle-\sqrt{p^2-2m\hbar\omega}\uparrow|\psi\rangle_0\Biggr].\cr}
\eqno(3.11)
$$
These two terms correspond to the transmitted and reflected components of the particle, whose spin state is flipped to the down state through the interaction with the potential.
The connection to the plane-wave solution is clear, since the common factor in the parentheses coincides with the transmission ($\cal T_\downarrow$) and reflection ($\cal R_\downarrow$) coefficients, which are both real and take the same form in this particular case.
Observe the presence of the $\theta$-function, which is necessary for energy conservation, since the total energy of the particle in the down state is given by $p^2/2m$ and the particle will gain energy $\hbar\omega$ by flipping its spin from up to down:
No spin-down state with energy less than $\hbar\omega$ is allowed because the incident particle has been prepared in the up state.

\bigskip
%------------------------------------------------------------------------------
\centerline{\bf\S4.~Estimation of the delay time}
%------------------------------------------------------------------------------
\medskip
     The exact solutions to the time-dependent Schr\"odinger equations we have obtained in the previous sections explicitly show that their wave-packet solutions, even though they possess energy-nonconserving components within the finite time duration, are to be confined to the energy shell in the asymptotic ($t\to\infty$) limit and converge to the states composed of the transmitted and reflected waves, multiplied by the relevant transmission ($\cal T$) and reflection ($\cal R$) coefficients for the plane wave solution, that is,
$$
\ppsi t\buildrel t\to\infty\over\longrightarrow 
e^{-iE_pt/\hbar}\left[{\cal T}(p)\ppsi0
+{\cal R}(p)\langle-p|\psi\rangle_0\right],
\eqno(4.1)
$$
where $E_p$ is the energy of the state $|p\rangle$.
For the moment, for notational simplicity, the dependence on the possible internal degrees of freedom of the particle shall be suppressed.

     In order to illustrate the asymptotic behavior of the wave packet more clearly and quantitatively, let us express the above wave function in ordinary configuration space.
We assume that the initial particle is represented by an almost monochromatic wave packet with a Gaussian profile
$$
\ppsi0
={\cal N}\exp\Bigl[-{(p-p_0)^2\over4\delta p^2}-{i\over\hbar}(p-p_0)x_0\Bigr],
\eqno(4.2)
$$
where $\cal N$ is a normalization constant and $\delta p\ll p_0$.
This wave packet is distributed around its mean position $x_0<0$ with width $\hbar/2\delta p$ and moves with an average momentum $p_0>0$ toward the potential.
In order not to make any sensible overlap between the initial wave packet and the potential located at the origin, the inequality $x_0+\hbar/2\delta p\ll0$ is also assumed.

      The transmitted wave packet in $x$-space is given by the Fourier transform of the first term of the RHS of Eq.~(4.1)
$$
\langle x|\psi\rangle_t^{\rm tr}
=\int{dp\over\sqrt{2\pi\hbar}}{\cal T}(p)e^{-iE_pt/\hbar+ipx/\hbar}\ppsi0.
\eqno(4.3)
$$
If the transmission coefficient ${\cal T}(p)$ is a slowly varying function of $p$, we may expand ${\cal T}(p)$ around $p_0$ as a power series of $p-p_0$ to perform a Gaussian integral over $p$ in (4.3).
Under this assumption, the transmitted wave packet (in the asymptotic region) turns out to take also the Gaussian form
$$
\langle x|\psi\rangle_t^{\rm tr}
\sim
{\cal T}(p_0)\exp\Bigl[
-{1\over4\hbar^2u}
\Bigl(x-x_0-p_0t/m-i\hbar{\cal T}'(p_0)/{\cal T}(p_0)\Bigr)^2
+{i\over\hbar}p_0x-{i\over\hbar}E_{p_0}t\Bigr],
\eqno(4.4)
$$
where
$$
u={1\over4\delta p^2}+{it\over2m\hbar}.
\eqno(4.5)
$$
It is straightforward, if lengthy, to extract the real part of the exponent in (4.4), which can be written in the following form
$$
-{\delta p^2\over\hbar^2}{1\over1+(4E_{\delta p}t/\hbar)^2}
\Bigl[x-x_0-v_{\rm tr}(t-\delta_{\rm tr})\Bigr]^2
+(\delta p^2/p_0^2)
 \Bigl[p_0{\rm Re}\Bigl({\cal T}'(p_0)/{\cal T}(p_0)\Bigr)\Bigr]^2,
\eqno(4.6)
$$
where $v_{\rm tr}$ stands for the mean velocity of the transmitted wave packet
$$
v_{\rm tr}
={p_0\over m}\Bigl[
  1+2(\delta p^2/p_0^2)p_0{\rm Re}\Bigl({\cal T}'(p_0)/{\cal T}(p_0)\Bigr)
 \Bigr].
\eqno(4.7)
$$
The quantity $\delta _{\rm tr}$ in (4.6), given by
$$
\delta _{\rm tr}
={m\hbar\over p_0}{\rm Im}\Bigl({\cal T}'(p_0)/{\cal T}(p_0)\Bigr)
 \Bigl[
  1+2(\delta p^2/p_0^2)p_0{\rm Re}\Bigl({\cal T}'(p_0)/{\cal T}(p_0)\Bigr)
 \Bigr]^{-1},
\eqno(4.8)
$$
measures a temporal displacement of the transmitted wave packet and is called the time delay or advance, according to its sign.
It is interesting to note that even though our derivation is quite different from that in [2], the main part of the above $\delta_{\rm tr}$ coincides with that obtained by taking the energy derivative of the scattering phase shift [2]:
The correction, however, is due to the wave packet effect, which is obtainable only when the wave packet itself is treated directly.

     Similarly, we shall define the reflected wave packet (in the asymptotic region) by
$$
\langle x|\psi\rangle_t^{\rm rf}
=\int{dp\over\sqrt{2\pi\hbar}}{\cal R}(p)e^{-iE_pt/\hbar+ipx/\hbar}
 \langle-p|\psi\rangle_0
\eqno(4.9)
$$
and perform a Gaussian integral over $p$, under similar conditions as in (4.4).
The behavior of the reflected wave packet, again shown to be a Gaussian, is read from the real part of the exponent
$$
-{\delta p^2\over\hbar^2}{1\over1+(4E_{\delta p}t/\hbar)^2}
\Bigl[x+x_0+v_{\rm rf}(t-\delta_{\rm rf})\Bigr]^2
+(\delta p^2/p_0^2)
 \Bigl[p_0{\rm Re}\Bigl({\cal R}'(p_0)/{\cal R}(p_0)\Bigr)\Bigr]^2,
\eqno(4.10)
$$
from which we obtain its mean velocity and the time delay
$$\eqalignno{
v_{\rm rf}
&={p_0\over m}\Bigl[
  1+2(\delta p^2/p_0^2)p_0{\rm Re}\Bigl({\cal R}'(p_0)/{\cal R}(p_0)\Bigr)
 \Bigr],&(4.11)\cr
\noalign{\smallskip}
\delta _{\rm rf}
&={m\hbar\over p_0}{\rm Im}\Bigl({\cal R}'(p_0)/{\cal R}(p_0)\Bigr)
 \Bigl[
  1+2(\delta p^2/p_0^2)p_0{\rm Re}\Bigl({\cal R}'(p_0)/{\cal R}(p_0)\Bigr)
 \Bigr]^{-1}.&(4.12)\cr}
$$

     It would be interesting to estimate the value of the delay times for simple potential scatterings we have considered in the previous sections.
Since the transmission and reflection coefficients for a particle with a definite momentum $p_0$, scattered by the $\delta$-potential given in (2.1), are given by
$$
{\cal T}(p_0)={1\over1+i\Omega(p_0)},\qquad
{\cal R}(p_0)={-i\Omega(p_0)\over1+i\Omega(p_0)},
\eqno(4.13)
$$
where $\Omega(p_0)\equiv mg/\hbar p_0$, the delay times for the transmitted and reflected wave packets are estimated to be
$$\eqalignno{
\delta_{\rm tr}
&={\hbar\over2}{\Omega(p_0)\over E_{p_0}}|{\cal T}(p_0)|^2
  \Bigl[1+2(\delta p^2/p_0^2)|{\cal R}(p_0)|^2\Bigr]^{-1},&(4.14)\cr
\noalign{\smallskip}
\delta_{\rm rf}
&={\hbar\over2}{\Omega(p_0)\over E_{p_0}}|{\cal T}(p_0)|^2
  \Bigl[1-2(\delta p^2/p_0^2)|{\cal T}(p_0)|^2\Bigr]^{-1}.&(4.15)\cr}
$$
Notice that both delay times $\delta_{\rm tr}$ and $\delta_{\rm rf}$ are positive, showing time delays of the wave packets, for a repulsive potential ($g>0$), while, for an attractive one ($g<0$), they become negative, which imply that the wave packets are actually advanced.
Observe also that these time delays vanish not only for the weak coupling limit $g\to0$, but also for the strong coupling limit $g\to\infty$.
Because in either limit, the wave packet is completely transmitted or reflected, we may understand that for the wave packet to exhibit a finite time delay or advance, the presence of both waves, transmitted and reflected, is crucial and that this phenomenon is ascribable to a kind of interference effect.
This is still at a speculative level, however, we can say that the time-delay phenomena have an essentially quantum origin and can not be compared with the classical counterpart, which would also be clear from the fact that the $\delta$-potential has a vanishing range of interaction.

      The nontrivial nature of the time delay may be seen in the other case considered in \S3., where a scattering of a spin-1/2 particle off the spin-flipping $\delta$-potential has been analyzed.
We have shown explicitly in (3.9) and (3.11) that the scattered wave packet in each channel is represented just like (4.1) with channel-dependent transmission and reflection coefficients.
If a particle prepared in the up state with definite momentum $p_0$ is scattered by the spin-flipping potential given in (3.1), these coefficients read
$$\eqalignno{
\left(\matrix{{\cal T}_\uparrow(p_0)  \cr
\noalign{\smallskip}
              {\cal T}_\downarrow(p_0)\cr}\right)
&={\sqrt{p_0^2}\over\sqrt{p_0^2(p_0^2+2m\hbar\omega)}+(mg/\hbar)^2}
  \left(\matrix{\sqrt{p_0^2+2m\hbar\omega}\cr
\noalign{\smallskip}
                -img/\hbar\cr}\right),&(4.16)\cr
\noalign{\smallskip}
\left(\matrix{{\cal R}_\uparrow(p_0)  \cr
\noalign{\smallskip}
              {\cal R}_\downarrow(p_0)\cr}\right)
&=-i{mg\over\hbar}{1\over\sqrt{p_0^2(p_0^2+2m\hbar\omega)}+(mg/\hbar)^2}
  \left(\matrix{-img/\hbar\cr
\noalign{\smallskip}
                \sqrt{p_0^2}\cr}\right).&(4.17)\cr}
$$
Notice that these quantities are essentially real and therefore no finite delay times are expected for any channels.
[See Eqs.~(4.8) and (4.12).]
This implies that no analogy to the previous simplest case without the spin-degrees of freedom is found!
Even though the appearance or disappearance of such delay times in potential scatterings has yet to be understood well, we can conclude that its analysis certainly requires completely quantum mechanical treatment.

\bigskip\bigskip
%------------------------------------------------------------------------------
\centerline{\bf Acknowledgements}
%------------------------------------------------------------------------------
\medskip
     The author would like to thank Profs.~M. Namiki and I. Ohba, Dr.~Y. Yamanaka and Saverio Pascazio for fruitful and helpful discussions.
This work is partially supported by Monbusho International Scientific Research Program:Joint Research (No.~08044097) and by Waseda University Grant for Special Research Projects No.~96A-126. 

\bigskip\bigskip
%------------------------------------------------------------------------------
\centerline{\bf References}
%------------------------------------------------------------------------------
\medskip
\item{[1]} W. Pauli, {\it Handbuch der Physik\/}, ed.~S. Fl\"ugge (Springer Verlag,1958) Vol.~5/1, p.~60.
\item{}    There have been many attempts to construct a time operator.
           See, for example, 
\item{}    T. Goto, K. Yamaguchi and N. Sudo, Prog.\ Theor.\ Phys.\ {\bf 66} (1981) 1525.
\item{}    T. Goto, S. Naka and K. Yamaguchi, Prog.\ Theor.\ Phys.\ {\bf 66} (1981) 1915.
\item{[2]} L. Eisenbud, dissertation, Princeton, 1948 (unpublished).
\item{}    E.P. Wigner, Phys.\ Rev.\ {\bf98} (1955) 145.
\item{}    M. Froissart, M.L. Goldberger and K.M. Watson, Phys.\ Rev.\ {\bf131} (1963) 2820.
\item{[3]} W. Elberfeld and M. Kleber, Am.\ J. Phys.\ {\bf56} (1988) 154.
\vfil\eject
%------------------------------------------------------------------------------
\centerline{\bf Figure captions}
%------------------------------------------------------------------------------
\bigskip
\item{}Fig.~1. Singularities of the integrand in (2.10) and the deformed contours for (a) $g>0$ and (b) $g<0$.
There are two simple poles, denoted by crosses, at $s_+=img^2/2\hbar^3$ and $s_-=-ip^2/2m\hbar$ and a cut, represented by a bold line, along the negative imaginary $s$-axis in either case.
Dashed lines run on the second Riemannian sheet.
\bigskip
\item{}Fig.~2. Deformed contours for the evaluation of (a) $\Cdown{t'}$ and (b) $\Cup{t'}$.
There are two simple poles, denoted by crosses, at $s'_+=i(\sqrt{\omega^2+(mg^2/\hbar^3)^2}-\omega)/2$ and $s'_-=-i(p^2/2m\hbar+\omega)$, two branch points, denoted by dots, at $s=0$ and $s=-i\omega$ and a cut, represented by a bold line, along the negative imaginary $s$-axis.
Dashed lines run on the second Riemannian sheet.
\bye